# Molecular dynamics simulations reveal the role of ceramicine B as novel PPARγ partial agonist against type 2 diabetes

Bidyut Mallick*

Galgotias College of Engineering and Technology, I, Knowledge Park-II, Greater Noida, Uttar Pradesh 201306, India

*Correspondence to:
Dr. Bidyut Mallick, PhD,
Department of Applied Science, Galgotias College of Engineering and Technology, Uttar Pradesh, India;

E-mail: bidyut.mallick@galgotiacollege.edu

# ABSTRACT

Peroxisome proliferator-activated receptors gamma (PPARγ) are ligand-activated controllers of various metabolic actions and insulin sensitivity. PPARγ is thus considered as an important target to treat type 2 diabetes. Available PPARγ drugs (full agonists) have robust insulin-sensitizing properties but are accompanied by severe side effects leading to complicated health problems. Here, we have used molecular docking and a molecular dynamics simulation study to find a novel PPARγ ligand from a natural product. Our study suggests that the inhibition of ceramicine B in the PPARγ ligand-binding domain (LBD) could act as a partial agonist and block cdk5-mediated phosphorylation. This result may provide an opportunity for the development of new anti-diabetic drugs by targeting PPARγ while avoiding the side effects associated with full agonists.

## INTRODUCTION

Peroxisome proliferator-activated receptor gamma (PPARγ), a nuclear receptor, is highly expressed in adipose tissue and controls hundreds of genes to regulate diverse biological functions, which includes glucose metabolism, lipid metabolism, and insulin sensitivity [1]. PPARγ dynamically shuttles between the nucleus and cytoplasm, but after binding with the agonists, it forms a heterodimer with the retinoid X receptor (RXR) and co-activators. This heterodimer is then translocated to the nucleus, where it regulates target genes involved in different metabolic functions and improves insulin sensitivity. Consequently, PPARγ has been considered as a suitable drug target for the treatment of type 2 diabetes. Thiazolidinediones (TZDs), such as pioglitazone and rosiglitazone, which are full agonists of PPARγ, are highly effective for this purpose, and they effectively lower blood glucose levels [2-5]. Full agonists bind in pocket I of the LBD of PPARγ and form hydrogen bonds with the Ser289, His323, His449 and Tyr473 residues, leading to a conformational change of H12 in the LBD [6]. This remodeling of the LBD allows PPARγ to dock with transcriptional coactivators and results in its activation [7]. Unfortunately, the use of TZDs is also associated with undesirable side effects, such as fat accumulation, fluid retention, cancer, loss of bone density and an increased risk of heart failure [8-11]. As a consequence, TZDs are no longer recommended for the treatment of type 2 diabetes. This ensures the need for new selective PPARγ ligands with improved clinical profiles.

Recent studies suggest that the obesity-induced CDK5-mediated phosphorylation of PPARγ at Ser245 (Ser273 for PPARγ2) results in the dysregulated expression of a series of genes associated with insulin resistance, which ultimately leads to diabetes [12, 13]. In the same study, it was also demonstrated that PPARγ ligands are capable of blocking this phosphorylation and are able to normalize the dysregulation of the PPARγ-target gene. These compounds (partial agonists) have a unique binding mode in the ligand-binding pocket of

PPARγ. Current findings show that partial agonists bind in the region between H3, the Ω-loop and the β-sheet, which form branch II of the PPARγ LBD, and that the region is different from the binding domain of a PPARγ full agonist. Most of the partial agonists interact via hydrogen bonding with Ser342 and via several hydrophobic interactions with the residues of branch II of the LBD of PPARγ, which causes the stabilization of the β-sheet/Ser245 surface and blocks phosphorylation. This mechanism exhibits antidiabetic effects similar to those of full agonists through the inhibition of PPARγ while ameliorating the side effects caused by the PPARγ transcriptional activity [12, 14, 15]. Until now, a number of PPAR$\gamma$ partial agonists have been reported, but none of them are currently FDA approved [16-21]. A few of these partial agonists have progressed through Phase II clinical trials, and several have not yet been tested for the later phases. Additionally, it has been reported that partial agonists exhibit a certain amount of transcriptional activation [21]. Thus, investigation on a new kind of partial agonist merits an intensive research focus.

Here, we have examined the agonist activity of ceramicines A–L (Figure 1) on PPAR$\gamma$ using molecular docking and a molecular dynamics simulation study. Ceramicine is similar in size to most of the other PPAR$\gamma$ partial agonists [19, 22], as it consists of a phenanthrene ring with a large variation of moieties at different positions, and both cyclopentane and furan ring systems (Figure 1). A recent experimental study suggests that ceramicine B acts as an anti-lipid droplets accumulation agent by interrupting the phosphorylation of FoxO1, which leads to the down regulation of the transactivation activity of PPARγ [22]; furthermore, studies on partial agonists reveal that the PPARγ activation by an H12-independent mechanism subsequently decreases the transcriptional activity of PPARγ [14, 15]. Examination of the binding affinity of ceramicine B and other ceramicines in the PPARγ ligand binding pocket would therefore be of great interest. Additionally, a large variation of moieties could allow for different binding possibilities in the LBD of PPARγ.

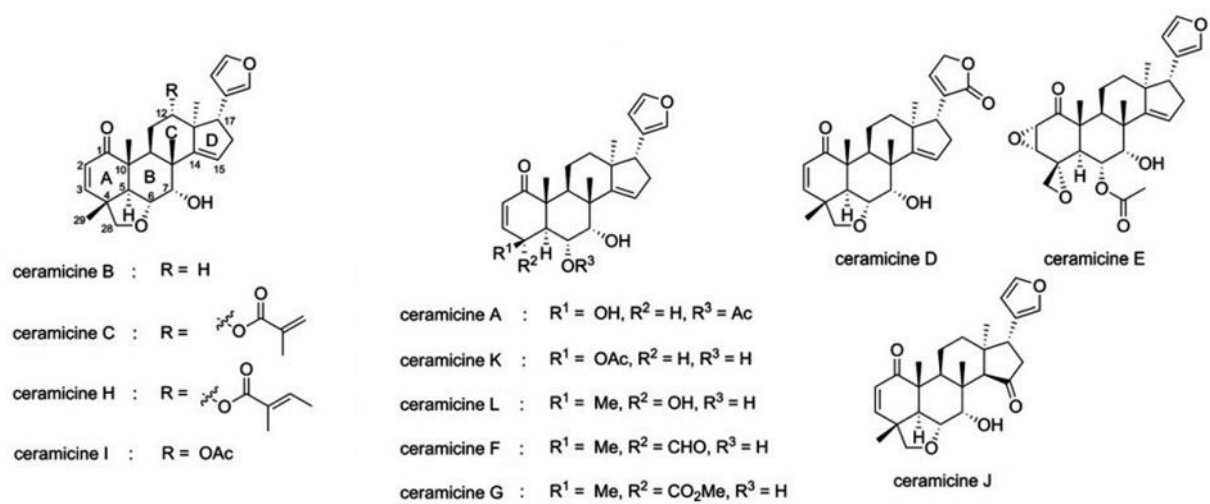

**Figure 1:** The molecular structure of ceramicine A-L.

## RESULTS AND DISCUSSION

**Molecular docking analysis**

In this study, we have examined the binding affinities of ceramicines A–L, Chelerythrine and SR1664 (Figure 1) using the AutoDock4 molecular docking program. Molecular docking is a computational method that performs the virtual screening of drug-like compounds based on target structures and identifies appropriate confirmations with a binding affinity score. This gives the fastest and most accurate prediction of drug-like candidates for a target protein and allows for further biological testing. Twelve ceramicine compounds (A to L) are known to originate from the bark of Malaysian C. ceramicus [23-26]. The docking of these 12 ceramicines shows four minimum energy poses in the LBD of PPARγ depending on the position of the moiety. Ceramicines B and D have similar lower binding energy conformations (docking pose 1, Figure 2A), with binding energies of -9.2 kcal/mol and -9.24 kcal/mol, respectively, filling branch II of the ligand binding pocket. The hydroxyl group at position 7 for both ceramicines B and D forms hydrogen bonds with the NH group of the Ser342 residue of the β-sheet, which is the key residue for partial agonist activation. These two molecules are also involved in several hydrophobic interactions with the residues of the H3, β-sheet and Ω-loop region (Figure 2A). Interactions of ceramicines B and D with PPARγ indicate that changes in the furan ring by furanone do not make a significant difference in the binding affinity of ceramicine.

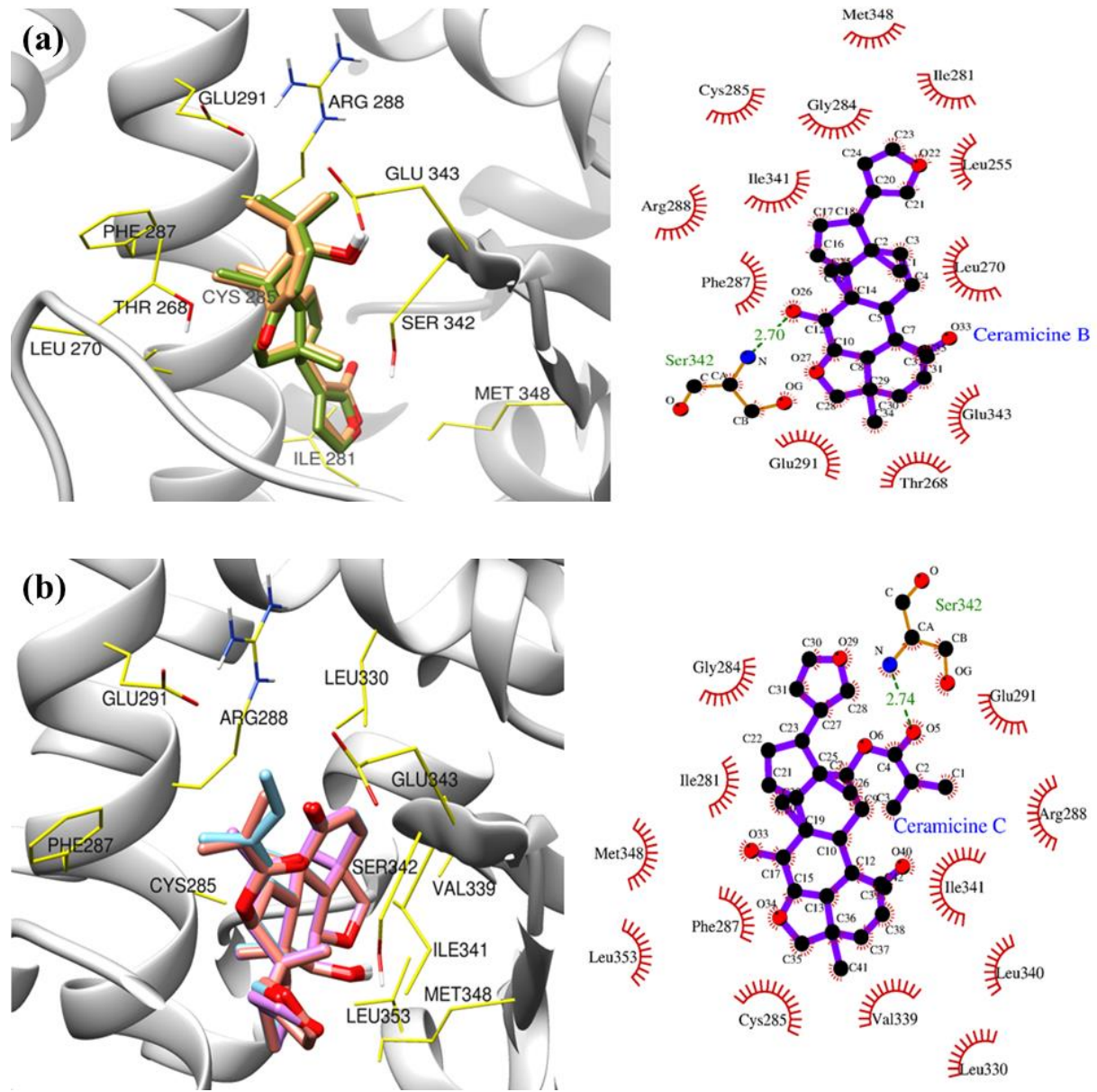

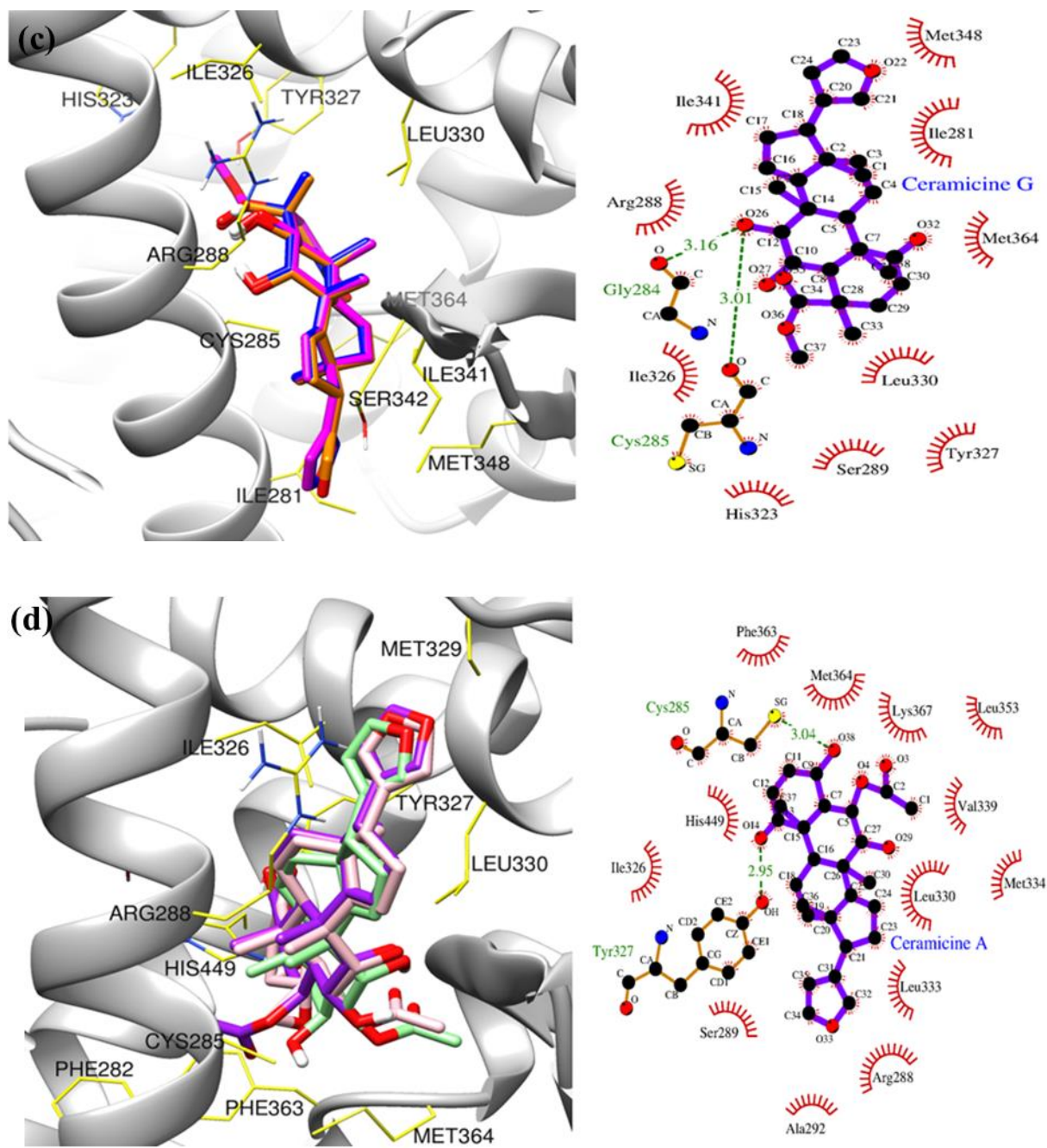

**Figure 2: Left:** The local structure of the docking complexes for ceramicine molecules in PPARγ LBD. (A) Ceramicine B and D (B) Ceramicine C, H and I (C) Ceramicine L, F and G (D) Ceramicine A, E and K. **Right:** Hydrophobic and hydrogen bond contacts between PPARγ protein and ceramicine compound determined by LIGPLOT program (A) Ceramicine B (B) Ceramicine C (C) Ceramicine G (D) Ceramicine A.

Ceramicines C, H and I have a methacrylate, a tiglate and an acetate moiety, respectively, at position 12. The substitution of an α-H at this position of ceramicine B by an ester group moiety leads to a conformational change in the ceramicine binding pose (docking pose 2, Figure 2B), as all these moieties form hydrogen bonds with the NH group of Ser342. The length of the hydrogen bond involved for ceramicines C, H and I are 2.74 Å, 2.75 Å and 2.82 Å, respectively. Hydrophobic contacts are formed with the residues of H3, H5, H6 and the β-sheet region, which allow the mentioned ceramicines to occupy branch II with the best binding conformations at energies of -9.53 kcal/mol, -9.89 kcal/mol, and -8.77 kcal/mol, respectively. Ceramicines B, C, D, H and I are all aligned parallel to the β-sheet in the LBD of PPARγ. While ceramicines B and D exhibit more hydrophobic interactions towards H3 and the Ω-loop, ceramicines C, H and I make hydrophobic contact with the residues of H5 and H6. None of these ceramicines are involved with the H12 residue in domain I of the binding pocket.

Substitutions by aldehyde, methoxycarbonyl, acetate and hydroxyl groups at positions 4 and 6 increase the polar surface in the phenanthrene ring of ceramicine and allow different binding conformations for ceramicines A, K, L, F and G in the binding pocket. Figure 2C illustrates the lowest energy docking poses of ceramicines L, F and G, which shows that the moiety in the phenanthrene ring interacts with the HIS323, TYR327, SER89, CYS285, ARG288 residues of branch I, maintaining hydrogen bonds with the CYS285 and GLY284 residues. The furan ring of these ceramicines reaches the β-sheet region of the LBD and forms hydrophobic interactions with the MET348, ILE341 and SER342 residues. The lowest binding energy conformations of ceramicines L, F and G have binding energies of -7.75 kcal/mol, -8.36 kcal/mol, and -8.7 kcal/mol, respectively. Conversely, the hydroxyl and acetate moiety at position R1 of ceramicines A and K forms hydrogen bonds with Cys285 and hydrophobic interactions with the residues of H11, H5, H7 and H3. The lowest energy docking pose spreads from the branch I to branch III of the LBD, as shown in Figure 2D. This is typical for the interaction of the PPARγ full agonist in the binding pocket. Ceramicine E (Figure 1) also shows the same binding conformations in the pocket. A similar conformation was again observed when cyclopentane was substituted by a cyclopentanone in ceramicine J.

The docking simulation study of the ceramicines in the LBD of PPARγ reveals that the position of the moiety in the phenanthrene ring, cyclopentane ring and furan ring is important in determining the binding conformation of ceramicine. Moiety variation in a particular position does not make much difference in the docking pose of the ligand; therefore, one ceramicine-PPARγ complex for each docking pose involved in partial agonist activity with the simplest ceramicine moiety and best binding affinity (ceramicine B, C and G) has been considered for further study.

We next evaluated the ADMET properties of ceramicine B, C and G by means of the Lipinski rule [27] using the FAFDrugs3 server [28]. The different parameters predicted were molecular weight, number of hydrogen bond donors and acceptors, topological PSA, number of rotatable bonds and log P. The ADMET parameters of the compounds are shown in Supplementary Table S1, which reveals that the values of the descriptors are within the optimum range followed by drug molecules. All three ceramicine compounds satisfied the Lipinski rule of five and were thus considered for the subsequent molecular dynamics simulation studies, which were performed to obtain the orientations of the compounds at the active site under dynamic conditions.

**Molecular dynamics simulation analysis**

A molecular dynamic study has been performed for the lowest energy conformations of ceramicines B, C and G. The best docking poses of these ceramicines in the docking simulation illustrate that they interact with the residues of branch II of the ligand binding pocket of the PPARγ target protein(PDB ID: 2F4B), which is similar to other partial agonists reported in previous studies [16-18]; however, the docking simulation only provides a static interaction mode of the protein-ligand complex. To investigate the protein-ligand interaction stability under dynamic conditions, a 100ns molecular dynamics simulation for the protein-ligand complexes has thus been performed and compared with the dynamic behavior of the uncomplexed PPARγ protein.

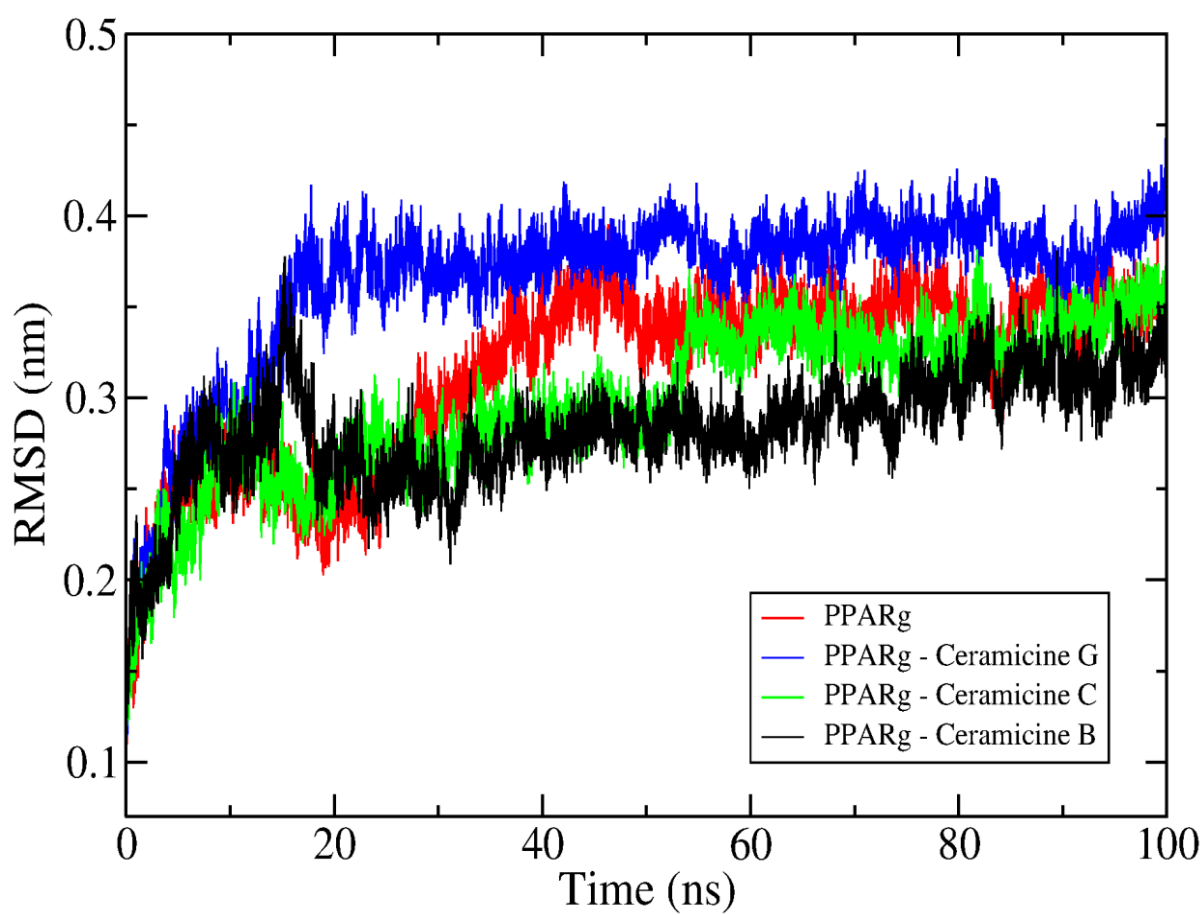

**Figure 3:** The RMSD value of the $C_\alpha$ atoms during 100 ns simulation of PPARγ(red), PPARγ -ceraicine G(blue), PPARγ -ceraicine C (green) and PPARγ -ceraicine B (black) system.

The RMSD is an important parameter used to differentiate the conformational stability of protein systems from MD simulation. The RMSD of $C_\alpha$ atoms ($C_\alpha$-RMSD), with respect to the structure present in the energy minimized, equilibrated system for each simulation, and has been calculated as a function of time. The plot in Figure 3 describes the $C_\alpha$-RMSD of the complexed and uncomplexed PPARγ protein during the 100-ns simulation. In Figure 3, it is clear that the $C_\alpha$-RMSD of all the PPARγ-ceramicine complexes equilibrated before 20 ns, whereas uncomplexed PPARγ took 40 ns to stabilize during simulation. The $C_\alpha$-RMSD of PPARγ bound with ceramicine G exhibited a constant deviation of ~0.38 nm after reaching the equilibrium state. In the case of the uncomplexed PPARγ and the PPARγ-ceramicine C complex, the RMSD equilibrated with slightly lower values of ~0.35 nm and 0.33 nm, respectively. The $C_\alpha$-RMSD for the PPARγ-ceramicine B complex had a smaller fluctuation than all the other considered systems. The PPARγ structure is therefore more stable when complexed with ceramicine B.

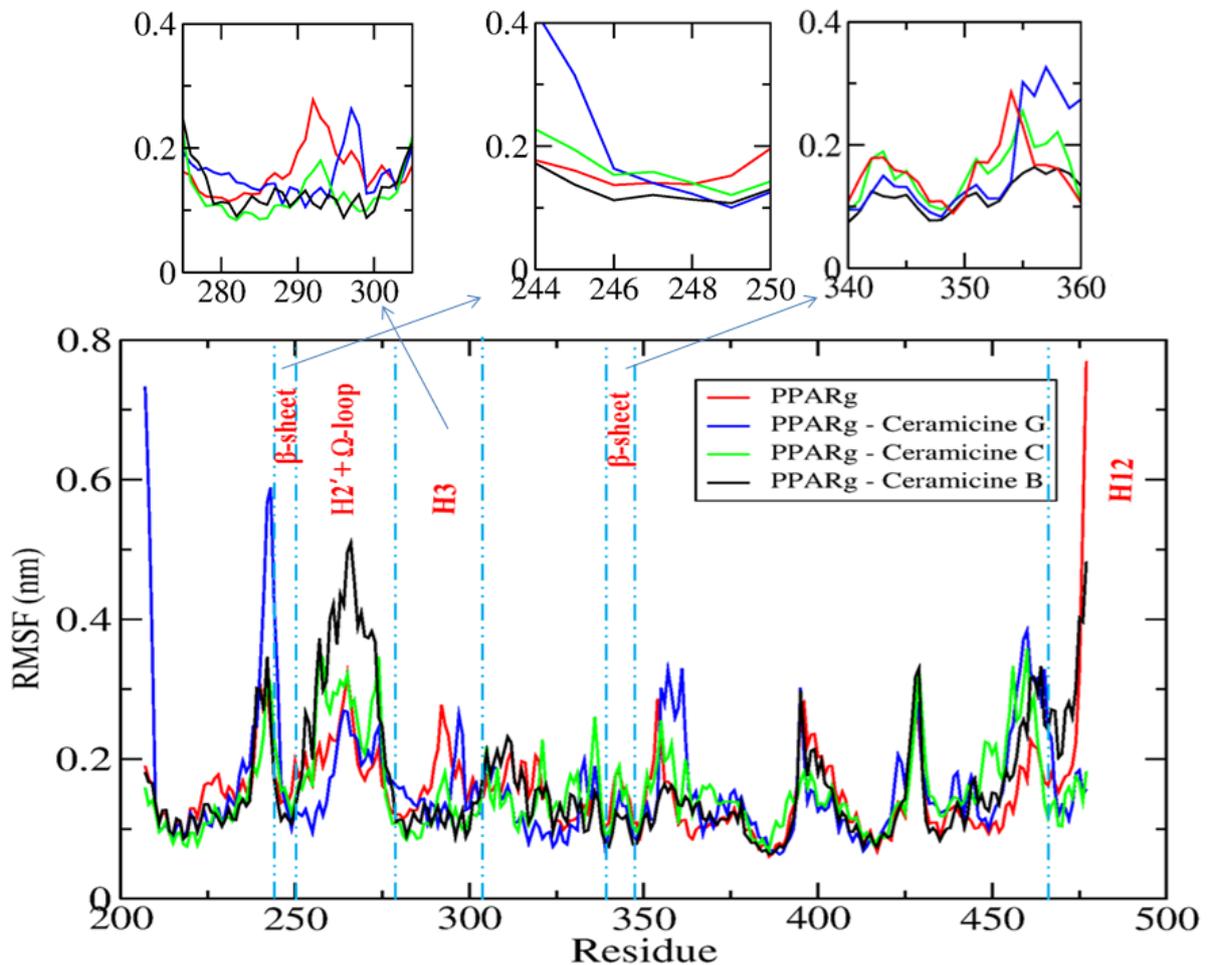

**Figure 4:** RMSF of $C_α$ atoms as a function of amino acids for PPARγ(red), PPARγ -ceraicine G(blue), PPARγ -ceraicine C (green) and PPARγ -ceraicine B (black) system.

To investigate the dynamics of the important residues in the complexes compared to the uncomplexed form, the root mean square fluctuations (RMSF) of the $rC_α$ atoms of the protein were calculated, as shown in Figure 4.The figure represents reduced RMSF values for H3 and the β-sheet when complexed with ceramicine B relative to other considered complexes. The displacement of the H12 residue of the PPARγ-ceramicine B complex is comparable to that of the uncomplexed form, whereas in the inhibition of ceramicine C, ceramicine G highly stabilizes helix 12; thus, RMSF variation shows that ceramicine B stabilizes the H3, β-sheet and Ser245 regions more strongly than the other ceramicines, which could indicate more efficiency in blocking the Cdk5-mediated phosphorylation of Ser245 [29]. Conversely, the dynamics of H12 remain unaffected due to the inhibition of ceramicine B, which reflects the transcriptional activity-independent inhibition of PPARγ [14, 15].Hence, ceramicine B could act as the most significant PPARγ partial agonist in comparison to other ceramicines.

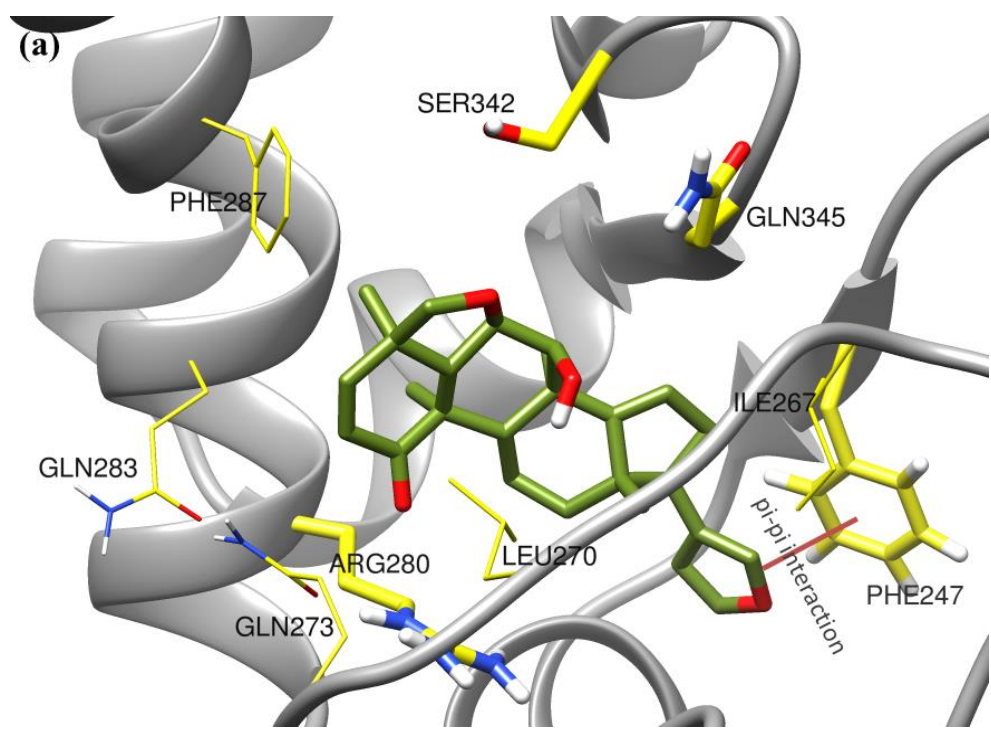

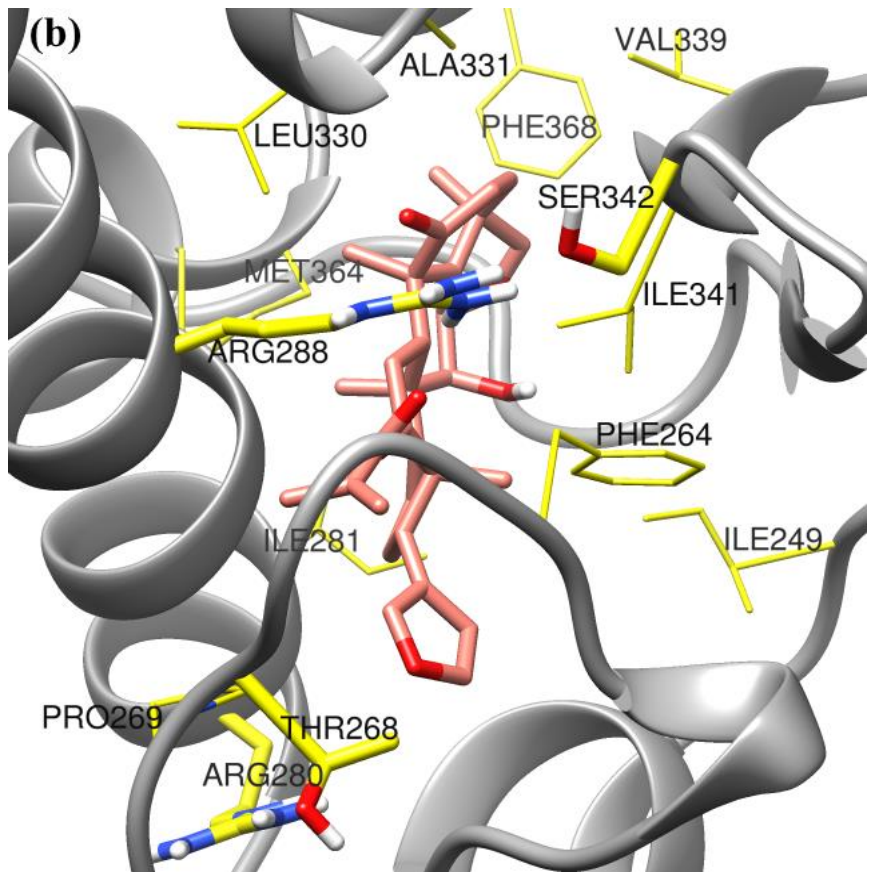

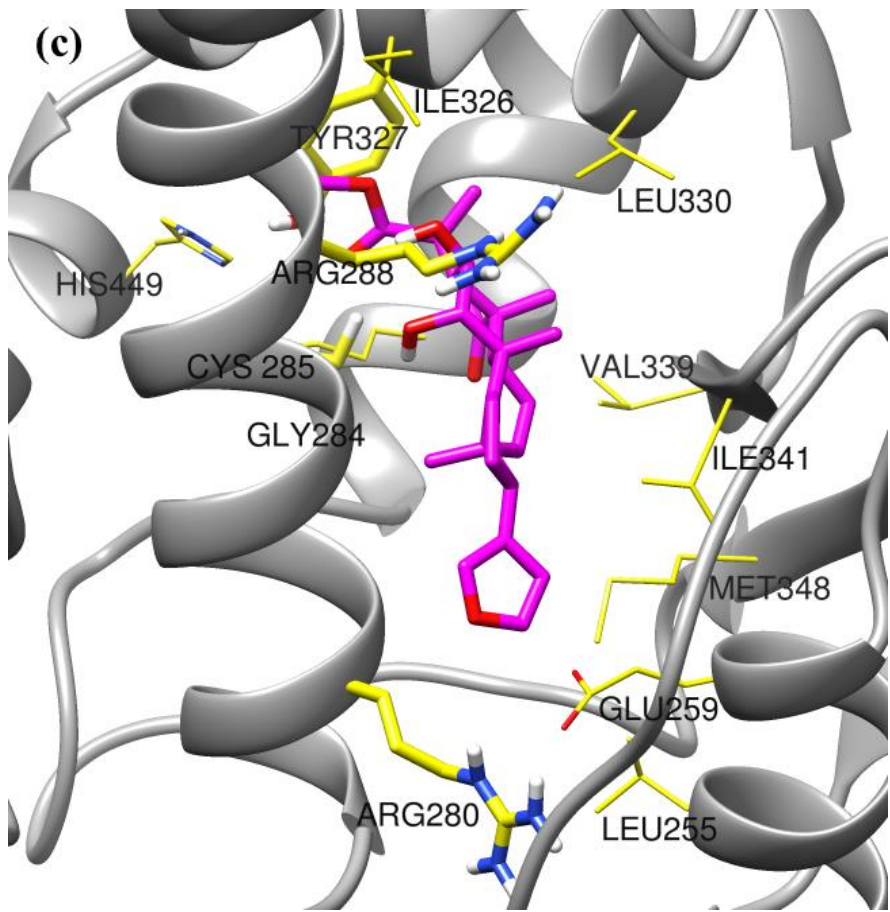

**Figure 5:** Final nanosecond snapshots of (A) Ceramicine B (B) Ceramicine C (C) Ceramicine G in PPARγ LBD obtained from molecular dynamics simulation. Residues with thick bond are making hydrogen bond interaction during the course of the simulation.

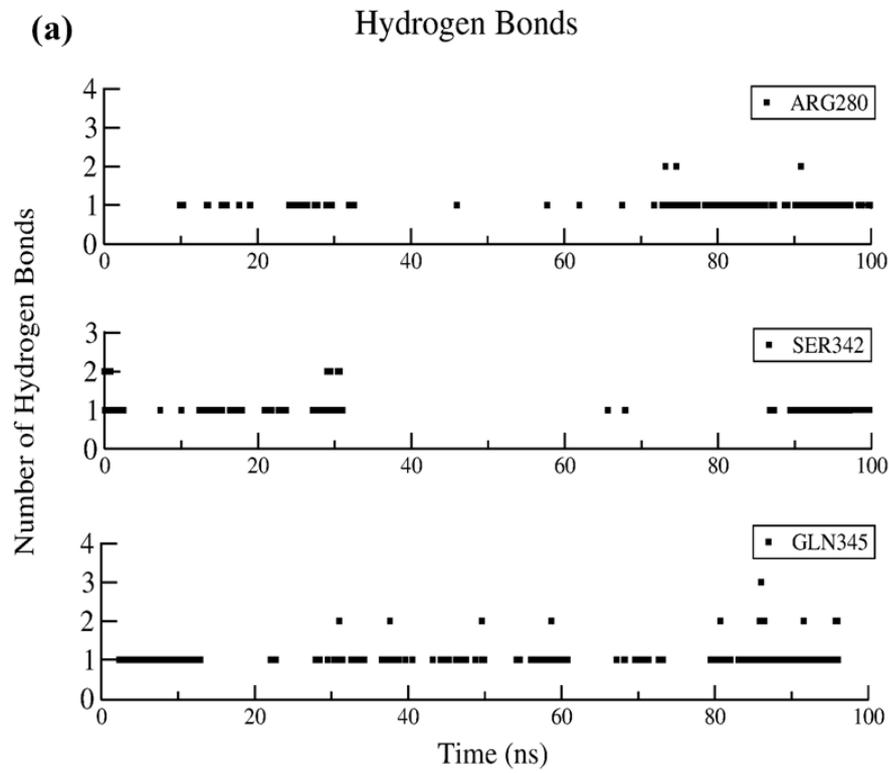

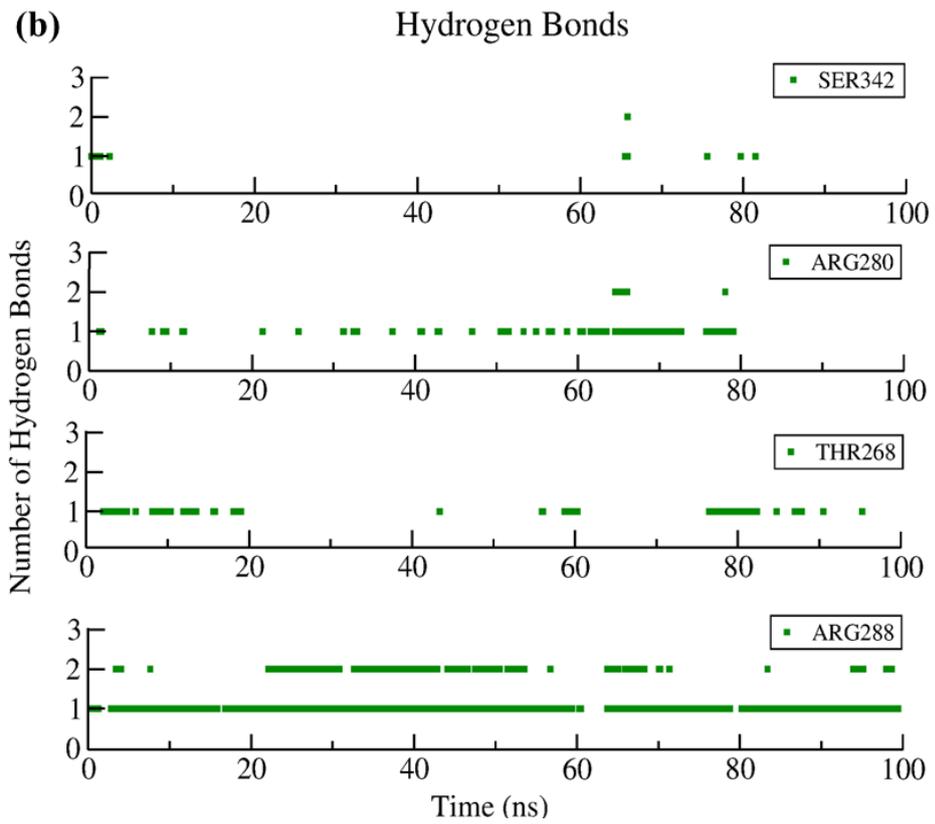

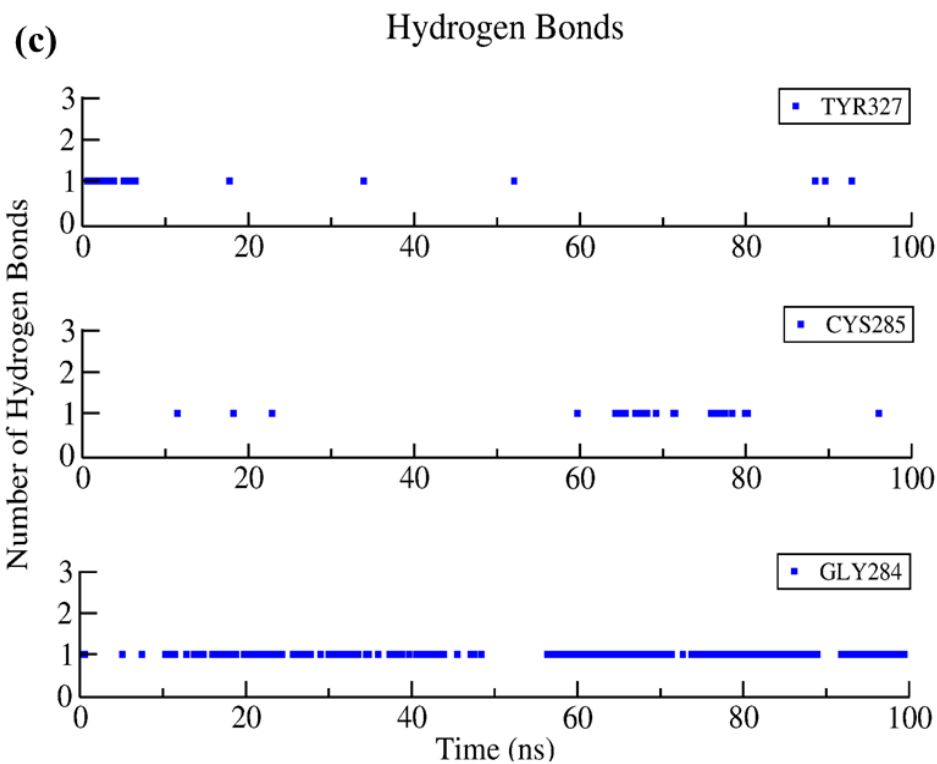

**Figure 6:** Number hydrogen bonds formed between residues of PPARγ and Ceramicine B (A), Ceramicine C (B), Ceramicne G (C) during the simulation.

To understand the interaction of ceramicines with PPARγ during the course of the simulation, the number of hydrogen bonds formed with different residues of the LBD has been calculated (Figure 6), and the pdb file at 100ns has been taken out to check the final confirmation of the inhibitor in the binding pocket. Figure 5 depicts the final binding modes of ceramicines B, C and G, which shows that all the ceramicines moved from their static docking poses during simulation due to the lack of constant hydrogen bond interactions with particular residues (Figure 6). Ceramicines C and G maintain their conformations, similar to other reported partial agonists [16, 17], and hydrogen bonding with Ser342 is significantly lower. Ceramicine C forms the maximum number of hydrogen bond contacts with Arg288, and ceramicine G forms the maximum number with Gly284. Ceramicine B, however, exhibits a different binding conformation previously not reported for any other partial agonist. After few nanoseconds of the simulation, ceramicine B moved to the position in (shown in Figure 5) which it forms hydrogen bonds with Ser342 and Gln345, and its furan ring forms a π–π stacking interaction with Phe247.Additionally, ceramicine B maintains a greater number of hydrogen bond contacts with Ser342 during the simulation in comparison to the other ceramicines. Thus, the new binding conformation of ceramicine B could be an effective PPARγ partial agonist, as it highly stabilizes the β-sheet/Ser245 region but does not affect the dynamics of H12.

## CONCLUSIONS

Our results demonstrate that different ceramicines from the bark of Malaysian *C. ceramicu* exhibit different binding conformations in the PPARγ LBD depending on the position of the moiety. Ceramicines have the shape and volume that fit into the region where partial agonists typically bind and moiety to bind with Ser342. This is particularly important because large ligands that occupy branch-II also spread towards branch-I and interact with helix 12. The docking simulation study shows that ceramicines B (D), C (H, I) and G (L, F) bind in a manner that could act as PPARγ partial agonists and that the lowest energy conformations of ceramicines B, C and G engage in H12-independent interactions. Further investigations using MD simulations suggest that ceramicine B constructs hydrogen bonds with the residues of the β-sheet and reduces the dynamics of the residues around the Ser245 region, keeping H12 unaffected, whereas other ceramicines altered the H12 dynamics with a lower number of hydrogen bond interactions with the β-sheet region. Thus, our finding shows a new binding conformation of an inhibitor (ceramicine B) in the LBD of PPARγ that significantly stabilizes the Ser245 region and could block Ser245 phosphorylation. A lower degree of H12 stabilization, which affects the recruitment of co-activators, decreases the transcriptional activity of PPARγ. Additionally, ceramicines show good ADMET properties that are within the optimum range of orally active drugs; furthermore, our results provide a new active molecule from natural extracts with antidiabetic properties that acts through the inhibition of Cdk5-mediated phosphorylation of PPARγ while lacking classical agonism.

## MATERIAL AND METHODS

**Data set**

The crystal structures of the PPARγ (isoform 1) receptor protein of Protein Data Bank (PDB) ID 2F4B [30] were downloaded from RCBS Protein Data Bank (http://www.pdb.org). The bound ligand and unwanted atoms were removed from the protein structure and minimized using the GROMACS-4.5.6 software package [31]. The 3D structures of ceramicines A–L were generated by using CORINA software, (https://www.molecular-networks.com/), which takes Isomeric SMILES from the Pubchem compound database

(http://www.ncbi.nlm.nih.gov/pccompound), and the structure was optimized with the help of Chimera 1.6.2.

**Molecular docking and virtual screening**

The docking study of the considered molecules into the PPARγ active site was performed with AutoDock4 [32]. AutoDock is one of the most widely used docking software programs that uses a Lamarckian genetic algorithm to predict the docking conformation between protein and ligand [33]. AutoDock Tools 1.5.6 was employed to prepare the protein and ligand and to perform the docking simulation. During the docking process, the protein was kept rigid and the ligands were flexible with all their torsional bonds free to rotate. A cubic grid box of size 60 Å×60Å×60 Å with points separated by 0.375 Å was generated and encompassed all the active site residues of PPARγ. All the default parameters were applied except the number of GA runs, which was set to 100. At the end of each docking, the lowest energy conformation was considered as the best binding conformation between each considered molecule and PPARγ. Cluster analysis showed that more than 20 percent of docking poses had the lowest energy conformations in each docking study.

The FAF Drugs3 web server [28] was used to analyze the ADMET properties of the best-docked ceramicine compounds obtained from the AutoDock result. ADMET properties play an extremely crucial role in the discovery and development of novel drugs at the earlier stage, as unfavorable ADMET properties have been identified as a major cause of failure even for the most promising drug candidate molecules [34]. While calculating these properties using the FAFDrugs3 server, all the filtering options were kept as default except the "logP computation program" and "physchem filters", which were set as "XLOGP3" and "Lipinski-RO5", respectively. According to this, the ligands were screened by Lipinski's rule of five, which suggests that compounds are within the acceptable range for a drug molecule if they possess a logP of less than 5, a molecular weight of less than 500, fewer than ten rotatable bonds and no more than 5 hydrogen-bond donors and ten hydrogen-bond acceptors.

**Molecular dynamic simulation**

The molecular dynamics simulations of the PPARγ protein and previously screened PPARγ-ceramicine complexes were performed using GROMACS-4.5.6 software [35] under the GROMOS96 53a6 force field [36, 37].The topology files of ceramicine compounds were generated using the online Automated Topology Builder (ATB) server, which applies quantum mechanical calculations combined with a knowledge-based approach to derive the GROMACS compatible force field parameters [38]. All the complexes were handled separately by putting them into a cubic box and were solvated by using the SPC216 water model. Na ions were added to each system to electrically neutralize the total charge by using the genion tool of the GROMACS package, which randomly substitutes water molecules with ions at the most favorable electrostatic potential positions. This was followed by energy minimization using a steepest decent algorithm with a maximum step size of 0.01 nm, maintaining a tolerance of 1000 kJ/mol/nm. The system was then subjected to equilibration at 300 K and 1 bar for 100ps under the conditions of position restraints for heavy atoms and LINCS constraints for all bonds [35]. Finally, the full system was subjected to a 100 ns MD simulation run, and the corresponding atom coordinates were stored every 0.002 ps during the simulation for later analyses.

**ABBREVIATIONS**

PPARγ: Peroxisome proliferator-activated receptors gamma; Ser: Serine; Cdk5: Cyclin-dependent kinase 5; Gly: Glycine; TZDs: Thiazolidinediones; LBD: ligand-binding domain; ADMET: Absorption, Distribution, Metabolism, Excretion and Toxicity; RCBS: Research Collaborators for structural Bioinformatics Protein; RMSD: Root-Mean-Square Deviation;

.

**ACKNOWLEDGMENTS**

This research was supported and funded by Galgotias Educational Institutions.

**CONFLICTS OF INTEREST**

The author declares that they have no competing interests.

**REFERENCES**

1. Tontonoz, P., & Spiegelman, B. M., (2008). Fat and beyond: the diverse biology of PPARγ. Annu Rev Biochem, 77, 289-312.

2. Lehmann, J. M., Moore, L. B., Smith-Oliver, T. A., Wilkison, W. O., Willson, T. M., & Kliewer, S. A., (1995). An antidiabetic thiazolidinedione is a high affinity ligand for peroxisome proliferator-activated receptor γ (PPARγ). Journal of Biological Chemistry, 270, 12953-6.

3. Willson, T. M., Lambert, M. H., & Kliewer, S. A., (2001). Peroxisome proliferator–activated receptor γ and metabolic disease. Annual review of biochemistry, 70, 341-67.

4. Elte, J., & Blickle, J., (2007). Thiazolidinediones for the treatment of type 2 diabetes. European journal of internal medicine, 18, 18-25.

5. Cariou, B., Charbonnel, B., & Staels, B., (2012). Thiazolidinediones and PPARγ agonists: time for a reassessment. Trends in Endocrinology & Metabolism, 23, 205-15.

6. Nolte, R. T., Wisely, G. B., Westin, S., Cobb, J. E., Lambert, M. H., Kurokawa, R., … Milburn, M. V., (1998). Ligand binding and co-activator assembly of the peroxisome proliferator-activated receptor-gamma. Nature. 395, 137-43.

7. Farce, A., Renault, N., & Chavatte, P., (2009). Structural insight into PPARγ ligands binding. Current medicinal chemistry. 16, 1768-89.

8. Nissen, S. E., & Wolski, K,. (2007). Effect of rosiglitazone on the risk of myocardial infarction and death from cardiovascular causes. N Engl j Med. 2007, 2457-71.

9. Rubenstrunk, A., Hanf, R., Hum, D. W., Fruchart, J-C., & Staels, B., (2007). Safety issues and prospects for future generations of PPAR modulators. Biochimica et Biophysica Acta (BBA)-Molecular and Cell Biology of Lipids, 1771, 1065-81.

10. Tang, W., & Maroo, A., (2007). PPARγ agonists: safety issues in heart failure. Diabetes, Obesity and Metabolism, 9, 447-54.

11. Kung, J., & Henry, R. R., (2012). Thiazolidinedione safety. Expert opinion on drug safety. 11, 565-79.

12. Choi, J. H., Banks, A. S., Estall, J. L., Kajimura, S., Boström. P., Laznik, D., … Blüher, M., (2010). Anti-diabetic drugs inhibit obesity-linked phosphorylation of PPARgamma by Cdk5. Nature, 466, 451-6.

13. Choi, J. H., Banks, A. S., Kamenecka, T. M., Busby, S. A., Chalmers, M. J., Kumar, N., Bruning, J. B., (2011). Anti-diabetic actions of a non-agonist PPARγ ligand blocking Cdk5-mediated phosphorylation. Nature, 477: 477.

14. Lu, I-L., Huang, C-F., Peng, Y-H., Lin, Y-T., Hsieh, H-P., Chen, C-T., … Prakash, E., (2006). Structure-based drug design of a novel family of PPARγ partial agonists: virtual screening, X-ray crystallography, and in vitro/in vivo biological activities. Journal of medicinal chemistry, 49, 2703-12.

15. Gelman, L., Feige, J. N., & Desvergne, B., (2007). Molecular basis of selective PPARγ modulation for the treatment of type 2 diabetes. Biochimica et Biophysica Acta (BBA)-Molecular and Cell Biology of Lipids. 1771, 1094-107.

16. Motani, A., Wang, Z., Weiszmann, J., McGee, L. R., Lee, G., Liu, Q., … Lindstrom, M., INT131: a selective modulator of PPARγ. Journal of molecular biology, 386, 1301-11.

17. Amato, A. A., Rajagopalan, S., Lin, J. Z., Carvalho, B. M., Figueira, A. C., Lu, J., … Souza, P. C.,(2012). GQ-16, a novel peroxisome proliferator-activated receptor γ (PPARγ) ligand, promotes insulin sensitization without weight gain. Journal of Biological Chemistry. 287, 28169-79.

18. Bruning, J. B., Chalmers, M. J., Prasad, S., Busby, S. A., Kamenecka, T. M., He, Y., … Griffin, P. R., (2007). Partial agonists activate PPARγ using a helix 12 independent mechanism. Structure. 15, 1258-71.

19. Zheng, W., Qiu, L., Wang, R., Feng, X., Han, Y., Zhu, Y.,… Li, Y., (2015). Selective targeting of PPARgamma by the natural product chelerythrine with a unique binding mode and improved antidiabetic potency. Sci Rep., 5, 12222.

20. de Groot, J. C., Weidner, C., Krausze, J., Kawamoto, K., Schroeder, F. C., Sauer, S., Büssow, K.,(2013). Structural characterization of amorfrutins bound to the peroxisome proliferator-activated receptor γ. Journal of medicinal chemistry, 56, 1535-43.

21. Kroker, A. J., & Bruning, J. B., (2015) Review of the structural and dynamic mechanisms of PPARγ partial agonism. PPAR research, 2015.

22. Wang, L., Waltenberger, B., Pferschy-Wenzig E-M., Blunder M., Liu, X., Malainer, C., … Heiss, E. H., (2014). Natural product agonists of peroxisome proliferator-activated receptor gamma (PPARγ): a review. Biochemical pharmacology, 92, 73-89.

23. Mohamad, K., Hirasawa, Y., Lim, C. S., Awang, K., Hadi, A. H. A., Takeya, K., Morita, H., (2008). Ceramicine A and walsogyne A, novel limonoids from two species of Meliaceae. Tetrahedron Letters, 49, 4276-8.

24. Mohamad, K., Hirasawa, Y., Litaudon, M., Awang, K., Hadi, A.H.A., Takeya, K., … Morita, H., (2009). Ceramicines B–D, new antiplasmodial limonoids from Chisocheton ceramicus. Bioorganic & medicinal chemistry, 17, 727-30.

25. Wong, C. P., Shimada, M., Nagakura, Y., Nugroho, A. E., Hirasawa, Y., Kaneda T, … Shiro, M., (2011). Ceramicines E—I, New Limonoids from Chisocheton ceramicus. Chemical and Pharmaceutical Bulletin. 59, 407-11.

26. Wong, C. P., Shimada, M., Nugroho, A. E., Hirasawa, Y., Kaneda, T., Hadi, A. H. A., … Morita, H., (2012). Ceramicines J–L, new limonoids from Chisocheton ceramicus. Journal of natural medicines. 2012; 66: 566-70.

27. Lipinski, C. A., Lombardo, F., Dominy, B. W., & Feeney, P. J., (2012). Experimental and computational approaches to estimate solubility and permeability in drug discovery and development settings. Advanced drug delivery reviews, 64, 4-17.

28. Lagorce, D., Sperandio, O., Baell, J. B., Miteva, M. A., & Villoutreix, B. O., (2015). FAF-Drugs3: a web server for compound property calculation and chemical library design. Nucleic acids research, 43, W200-W7.

29. Hughes, T. S., Chalmers, M. J., Novick, S., Kuruvilla, D. S., Chang, M. R., Kamenecka, … Griffin, P.R., (2012). Ligand and receptor dynamics contribute to the mechanism of graded PPARγ agonism. Structure. 20, 139-50.

30. Mahindroo, N., Wang, C-C., Liao, C-C., Huang, C-F., Lu, I-L., Lien, T-W., …, Hsu, M-C., (2006). Indol-1-yl acetic acids as peroxisome proliferator-activated receptor agonists: design, synthesis, structural biology, and molecular docking studies. Journal of medicinal chemistry, 49, 1212-6.

31. Hess, B., Kutzner, C., Van Der Spoel, D., & Lindahl, E.,(2008). GROMACS 4: algorithms for highly efficient, load-balanced, and scalable molecular simulation. Journal of chemical theory and computation, 4, 435-47.

32. Morris, G. M., Huey, R., Lindstrom, W., Sanner, M. F., Belew, R. K., Goodsell, D. S., Olson, A. J.,(2009). AutoDock4 and AutoDockTools4: Automated docking with selective receptor flexibility. Journal of computational chemistry, 30, 2785-91.

33. Morris, G. M., Goodsell, D. S., Halliday, R. S., Huey, R., Hart, W. E., Belew, R. K., Olson, A. J., (1998). Automated docking using a Lamarckian genetic algorithm and an empirical binding free energy function. Journal of computational chemistry, 19, 1639-62.

34. Yamashita, F., & Hashida, M.,(2004). In silico approaches for predicting ADME properties of drugs. Drug metabolism and pharmacokinetics, 19, 327-38.

35. Hess, B., Bekker, H., Berendsen, H. J., Fraaije, J. G.,(1997). LINCS: a linear constraint solver for molecular simulations. Journal of computational chemistry, 18, 1463-72.

36. van Gunsteren, W. F., Billeter, S. R., Eising, A. A., Hünenberger, P. H., Krüger, P., Mark, A. E., … Tironi, I. G., (1996). Biomolecular simulation: the {GROMOS96} manual and user guide.

37. Oostenbrink, C., Villa, A., Mark, A. E., & Van Gunsteren, W. F., (2004). A biomolecular force field based on the free enthalpy of hydration and solvation: the GROMOS force-field parameter sets 53A5 and 53A6. Journal of computational chemistry, 25, 1656-76.

38. Malde, A. K., Zuo, L., Breeze, M., Stroet, M., Poger, D., Nair, P. C., Oostenbrink, C., & Mark, A. E., (2011). An automated force field topology builder (ATB) and repository: version 1.0. Journal of chemical theory and computation, 7, 4026-37.

**SUPPLEMENTARY FILE**

Table1. ADMET properties of ceramicine B, C and G as predicted by FAFDrugs3 server (Lagorce et al., 2015).

| Parameters | Ceramicine B | Ceramicine C | Ceramicine G | Chelerythrine |
|---|---|---|---|---|
| **Molecular Weight (g/mol)** | 408.53 | 492.60 | 454.56 | 348.37 |
| **HBD** | 1 | 1 | 2 | 0 |
| **HBA** | 4 | 6 | 6 | 5 |
| **Topological PSA (Å$^2$)** | 59.67 | 85.97 | 96.97 | 40.80 |
| **Rotatable Bonds** | 1 | 4 | 3 | 2 |
| **logP** | 3.94 | 4.14 | 3.22 | 4.58 |
| **Lipinski Violations** | 0 | 0 | 0 | 0 |